\documentclass{article}
\usepackage{FileSettings/arxiv}
\usepackage[utf8]{inputenc} 
\usepackage[T1]{fontenc}    
\usepackage{url}            
\usepackage{booktabs}       
\usepackage{amsfonts}       
\usepackage{amsmath}
\usepackage{nicefrac}       
\usepackage{microtype}      
\usepackage{lipsum}		
\usepackage{graphicx}
\usepackage{listings}
\usepackage[table]{xcolor}
\usepackage[pdftex,colorlinks=true, allcolors=blue, linkcolor=blue]{hyperref}
\usepackage{xcolor}
\usepackage[most]{tcolorbox} 
\usepackage[framemethod=tikz]{mdframed}
\usepackage[labelfont=bf]{caption}
\usepackage{hyperref}
\usepackage{tabularx}

\renewcommand{\headeright}{Manual}
\renewcommand{\undertitle}{Manual}
\renewcommand{\shorttitle}{\textit{PRCpy}}

\definecolor{codegreen}{rgb}{0,0.6,0}
\definecolor{codegray}{rgb}{0.5,0.5,0.5}
\definecolor{codepurple}{rgb}{0.58,0,0.82}
\definecolor{backcolour}{rgb}{0.95,0.95,0.92}

\lstdefinestyle{mystyle}{
  commentstyle=\color{codegreen},
  keywordstyle=\color{magenta},
  numberstyle=\tiny\color{codegray},
  stringstyle=\color{codepurple},
  basicstyle=\ttfamily\footnotesize,
  breakatwhitespace=false,         
  breaklines=true,                 
  captionpos=b,                    
  keepspaces=true,                 
  numbers=left,                    
  numbersep=10pt,                  
  showspaces=false,                
  showstringspaces=false,
  showtabs=false,                  
  tabsize=2,
  xleftmargin=5pt,
  aboveskip=5pt,
  belowskip=0pt
}

\definecolor{light-gray}{gray}{0.95}

\surroundwithmdframed[
  hidealllines=true,
  backgroundcolor=light-gray,
  innerleftmargin=0pt,
  skipabove=10pt,
  skipbelow=20pt]{lstlisting}

\definecolor{dgreen}{RGB}{30, 143, 60}
\definecolor{vermilion}{RGB}{213, 94, 0}
\definecolor{hnavy}{RGB}{84, 120, 247}
\definecolor{hred}{RGB}{237, 43, 17}
\definecolor{hgreen}{RGB}{66, 220, 18}
\definecolor{dpink}{RGB}{236, 66, 245}

\definecolor{HeaderGray}{gray}{0.85}	
\definecolor{Gray}{gray}{0.9}
\definecolor{LightCyan}{rgb}{0.78,0.97,0.97}

\newcommand{\subfigref}[2]{\hyperref[{#1}]{\ref*{#1}#2}}
\input{FileSettings/BibSettings}

\begin{document}

\title{\emph{PRCpy}: A python package for processing of physical reservoir computing}

\author{\href{https://orcid.org/0009-0007-3004-6367}{
        \includegraphics[scale=0.06]{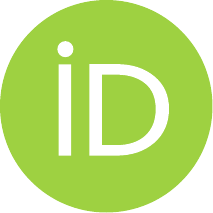}\hspace{1mm}\textbf{Harry Youel}}\textsuperscript{1,2,*} \and
\href{https://orcid.org/0009-0004-6156-2610}{
        \includegraphics[scale=0.06]{Figures/Misc/orcid.pdf}\hspace{1mm}\textbf{Daniel Prestwood}}\textsuperscript{\textbf{1,3,4,**}} \and
    \href{https://orcid.org/0000-0002-3469-0850}{
        \includegraphics[scale=0.06]{Figures/Misc/orcid.pdf}\hspace{1mm}Oscar Lee}\textsuperscript{1} \and
    \href{https://orcid.org/0009-0000-5837-7466}{\includegraphics[scale=0.06]{Figures/Misc/orcid.pdf}
        \hspace{1mm}Tianyi Wei}\textsuperscript{1,3} \and
    \href{https://orcid.org/0000-0003-0955-3640}{
        \includegraphics[scale=0.06]{Figures/Misc/orcid.pdf}\hspace{1mm}Kilian D. Stenning}\textsuperscript{4} \and
    \href{https://orcid.org/0000-0002-7044-7399}{
        \includegraphics[scale=0.06]{Figures/Misc/orcid.pdf}\hspace{1mm}Jack C. Gartside}\textsuperscript{4} \and
    \href{https://orcid.org/0000-0002-4821-4097}{
        \includegraphics[scale=0.06]{Figures/Misc/orcid.pdf}\hspace{1mm}Will R. Branford}\textsuperscript{4} \and
    \href{https://orcid.org/0000-0001-8767-6633}{
        \includegraphics[scale=0.06]{Figures/Misc/orcid.pdf}\hspace{1mm}Karin Everschor-Sitte}\textsuperscript{5} \and
    \href{https://orcid.org/0000-0002-2021-1556}{
        \includegraphics[scale=0.06]{Figures/Misc/orcid.pdf}\hspace{1mm}Hidekazu Kurebayashi}\textsuperscript{1,3,6,***}
}

\date{}

\maketitle

\begin{center}
    \textsuperscript{1}London Centre for Nanotechnology, University College London,\\ London, WC1H 0AH, United Kingdom \\

    \textsuperscript{2}Department of Physics and Astronomy, University College London,\\ London, WC1E 6BT, United Kingdom\\

    \textsuperscript{3}Department of Electronic and Electrical Engineering, University College London,\\ London, WC1E 7JE, United Kingdom\\

    \textsuperscript{4}Blackett Laboratory, Imperial College London,\\ London SW7 2AZ, United Kingdom\\
    
    \textsuperscript{5}Faculty of Physics and Center for Nanointegration Duisburg-Essen (CENIDE), University of Duisburg-Essen, Lotharstraße 1, 47057 Duisburg, Germany\\

    \textsuperscript{6}WPI Advanced Institute for Materials Research, Tohoku University, Sendai, Japan

    \vspace{0.25cm}\textsuperscript{*}\texttt{harry.youel.19@ucl.ac.uk},\textsuperscript{**}\texttt{daniel.prestwood.22@ucl.ac.uk},\textsuperscript{***}\texttt{h.kurebayashi@ucl.ac.uk} \\
    
\end{center}

\vspace{1cm}

\begin{abstract}
Physical reservoir computing (PRC) is a computing framework that harnesses the intrinsic dynamics of physical systems for computation. It offers a promising energy-efficient alternative to traditional von Neumann computing for certain tasks, particularly those demanding both memory and nonlinearity. As PRC is implemented across a broad variety of physical systems, the need increases for standardised tools for data processing and model training. In this manuscript, we introduce \textit{PRCpy}, an open-source Python library designed to simplify the implementation and assessment of PRC for researchers. The package provides a high-level interface for data handling, preprocessing, model training, and evaluation. Key concepts are described and accompanied by experimental data on two  benchmark problems: nonlinear transformation and future forecasting of chaotic signals. Throughout this manuscript, which will be updated as a rolling release, we aim to facilitate researchers from diverse disciplines to prioritise evaluating the computational benefits of the physical properties of their systems by simplifying data processing, model training and evaluation.
\end{abstract}
\keywords{Reservoir computing  \and RC \and Physical reservoir computing \and PRC}
\section{Introduction}

As machine learning and artificial intelligence (AI) develop in complexity, so does the demands for vast quantities of data and accompanying processing power ~\cite{openai2023gpt4,rombach2022highresolution,gozalobrizuela2023chatgpt}. Conventional computer hardware increasingly struggles to keep pace with the growing requirements of AI, with huge energy footprints due to an intrinsic mismatch between the hardware architecture and the demands of the software algorithms. The root causes of such energy inefficiencies arise in large part from the discrete memory-processor architecture, known as the von Neumann bottleneck. Consequently, its environmental carbon impact has come under increased scrutiny, directly impacting the cost and scaling of AI developments~\cite{dhar2020carbon,FREITAG2021100340,openaiCompute,strubell2019energy,strubell2020energy,schwartz2020green}.

The field of unconventional computing is broad and includes different frameworks that solve specific problems. Examples include neuromorphic computing \cite{schuman2022opportunities,stenning2023neuromorphic,marrows2024neuro}, probabilistic computing \cite{borders2019smtjs,chowdhury2023review,camsari2017p_bits}, and in-memory computing \cite{bao2022inmem,zhou2024inmem}. A particular advantage is that physical systems from diverse disciplines can offer distinct computational functionality via their intrinsic physical dynamics~\cite{roy_NaturePerspective2019,schuman2022opportunities,joksas2022memristive}. Among these, a neuromorphic computing framework termed \textit{physical reservoir computing} (PRC) has received attention thanks to its relatively simple implementation and potential for efficiently processing dynamic data. This is achieved by offloading nonlinear and memory-dependent processing to complex physical systems, leaving only a computationally-light linear final layer to train in software~\cite{yan2024emerging,liang2024physical,tanaka2019recent,nakajima2021reservoir}. As with other unconventional computing frameworks, examples of PRC have been demonstrated across a wide range of disciplines such as, magnetism~\cite{lee2024task,Torrejon2017,Gartside2022,Allwood_2023_APL,Vidamour2023,yokouchi2022pattern,korber2023pattern,furuta2018macromagnetic,edwards2023passive,taniguchi2022spintronic,papp2021characterization,pinna2020reservoir,watt2020reservoir,nomura2021reservoir,markovic2019reservoir,lee2022reservoir}, optics\cite{VanderSandeBrunnerSoriano+2017+561+576,PhysRevX.7.011015,PhysRevApplied.20.014051,PhysRevX.10.041037,Vandoorne2014,ng2024retinomorphicmachinevisionnetwork}, bio-materials~\cite{4218885,10.1371/journal.pcbi.0020165}, analogue electronics~\cite{Milano2022,Du2017,Moon2019}, and others~\cite{OBST2013189,Fujita_Yonekura_Nishikawa_Niiyama_Kuniyoshi_2018,Hauser2012}.

While many publications provide an excellent introduction to the general background theory and the working mechanisms of the PRC scheme, its detailed practical implementation methods are often not discussed at length. Therefore, providing such a technical discussion with practical examples may help circumvent barriers to entry in the field. Such practical details include processing the physical readout data, creating a reservoir matrix, training the linear readout layer of the model, and quantitatively evaluating system performance. While software-based reservoir computing (RC) solutions exist~\cite{trouvain2020reservoirpy,miao2024quantumreservoirpy,cabessa2022esntorch}, to our knowledge none currently focuses specifically on PRC. Here, we provide an open-source Python library, \textit{PRCpy} to address this need.

This manuscript is organised into four parts. Section~\ref{RC_background} outlines the fundamental background of PRC. Section~\ref{PRCpy_RC} provides a hands-on description of the setup and use of \textit{PRCpy} for implementing PRC. In Sec.~\ref{PRCpy_examples} we demonstrate nonlinear signal transformation and forecasting chaotic time-series data in two simple electronic circuits and in a complex magnetic system~\cite{lee2023perspective,lee2024task}. Finally, Sec.~\ref{conclusion} provides the conclusion and outlook. 
\section{Reservoir computing}
\label{RC_background}

\subsection{Working principles of reservoir computing}

Reservoir computing is an unconventional computing methodology that implements recurrent neural networks (RNNs). The scheme originated from a combination of two approaches, termed `echo state networks' (ESN) by Jaeger~\cite{jaeger2001echo} and `liquid-state machines' by Maass et al.~\cite{Maass_NeuralComm2002}. The core concept of RC involves constructing a network by initialising a set of nonlinear nodes with randomised internal recurrent connectivity and randomised weights at each node. This internal structure is typically termed the `reservoir', and in physical reservoir computing, this is represented by the complex physical system. During training, the internal structure is left fixed in its initial randomised state. Only a simple linear layer is trained (typically linear, ridge or logistic regression) - posing an alternative to conventional neural network architectures where every network weight is optimised during training, often through backpropagation\cite{jaeger2001echo,Maass_NeuralComm2002,Steil_IEEE2004,nakajima2021reservoir,Nakajima_2020}. This simple linear training is especially attractive for time-domain problems demanding recurrency such as future prediction, where the backpropagation training procedure for non-reservoir recurrent neural networks is highly computationally expensive. 

The complex internal structure of the reservoir nonlinearly transforms and `projects' data into a high-dimensional output space. Due to the internal recurrency/memory, nonlinear activation and dimensional expansion provided by the reservoir, tasks that were linearly inseparable in the input space now become linearly solvable in the high-dimensional output space. As only the final linear output layer is trained, RC typically offers computationally light training. In some cases, the linear output layer of the reservoir is replaced by a multi-layer perceptron style network of smaller size and complexity than the larger internal reservoir structure. 

\begin{figure}
\centering
\includegraphics[width=0.8\linewidth]{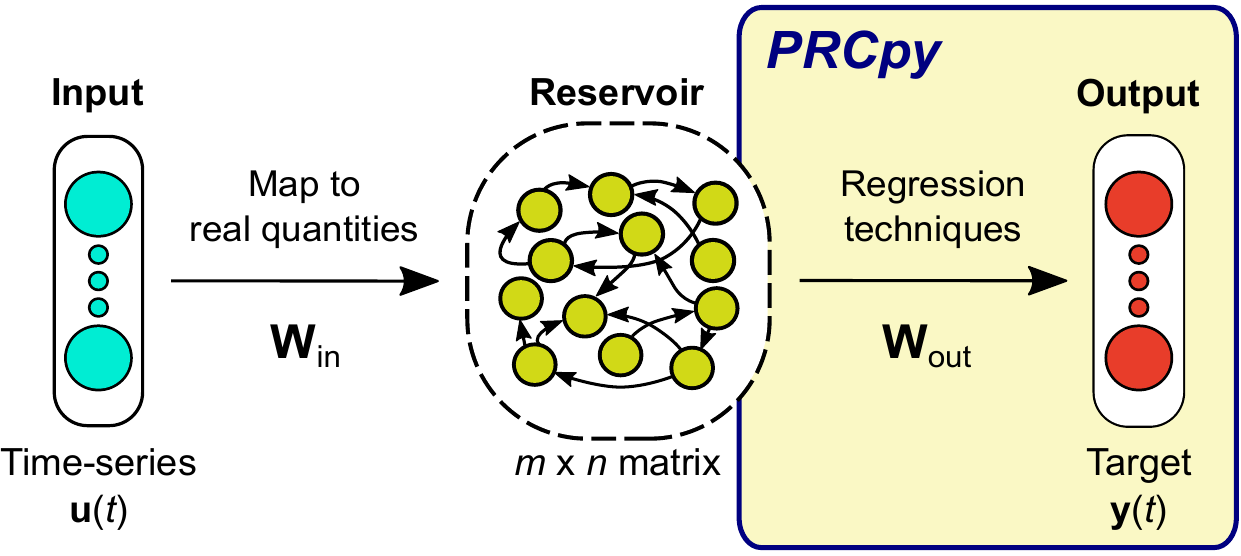}
\caption{\textbf{Typical PRC schematic.} The input layer consists of a time series signal. Each value is mapped onto a controllable quantity and applied to the physical system. The reservoir is a multidimensional matrix constructed by combining all system measurement responses. The elements of the reservoir are then used to predict the desired target output via regression techniques, often computed by an external computing device and program code. \textit{PRCpy} streamlines such complex coding required for the output layer.}
\label{Fig1_RC_overview}
\end{figure}

Figure~\ref{Fig1_RC_overview} depicts a typical (P)RC scheme involving three layers: the input, the reservoir, and the output. The reservoir computing framework only requires training model weights at the output layer, as mentioned above. The following provides a breakdown of each layer:

\begin{enumerate}
\item \textbf{Input Layer}: A time series $\mathbf{u}(t)$ is input to the reservoir via a weight matrix $\mathbf{W}_{\text{in}}$~\cite{Steil_IEEE2004,nakajima2021reservoir,jaeger2001echo,Maass_NeuralComm2002}. The result of this transformation, $\mathbf{x}_{\text{in}}(t)$ can be expressed as:

\begin{equation}
\mathbf{x}_{\text{in}}(t) = \mathbf{W}_{\text{in}} \cdot \mathbf{u}(t)
\end{equation}

\item \textbf{Reservoir}: The reservoir is typically a dynamic RNN that stores information about both current and past inputs in its internal states $\mathbf{x}(t)$, which evolve over time using fixed, randomly initialized internal weights $\mathbf{W}_{\text{res}}$. The state $\mathbf{x}(t)$ at time $t$ is a function of the previous states and the current input, described by:

\begin{equation}
\mathbf{x}(t) = f(\mathbf{W}_{\text{res}}(t-n) + \mathbf{x}_{\text{in}}(t))
\end{equation}

where $f$ is a nonlinear function, such as a hyperbolic tangent or sigmoid function in software-based reservoirs and $n$ represents an arbitrary number of timesteps into the past. In PRC, the internal physical dynamics of a material system serve as the reservoir. Crucially, while the nonlinear activation functions in software reservoirs are typically uniform across all internal reservoir nodes, in physical reservoirs the internal reservoir structure can contain a broad variety of history-dependent nonlinear activations provided by the range of physical dynamics active in the system.

\item \textbf{Readout Layer}: A subset of the internal reservoir nodes are defined as output nodes $\mathbf{x}_{\text{out}}(t)$. The readout layer linearly multiplies each reservoir output node by a weight value $\mathbf{W}_{\text{out}}$ optimised via a linear regression process, then sums these values to produce the output $\mathbf{y}(t)$. The reservoir output is given by:

\begin{equation}
\mathbf{y}(t) = \mathbf{W}_{\text{out}} \cdot \mathbf{x}_{\text{out}}(t)
\end{equation}
\end{enumerate}

This mapping technique enables spatio-temporal feature selection at the readout layer and only requires training $\mathbf{W}_{\text{out}}$. As mentioned above, the main advantage of this concept is the low training cost, as it only necessitates computationally inexpensive linear regression techniques~\cite{Steil_IEEE2004,nakajima2021reservoir,jaeger2001echo,Maass_NeuralComm2002,Jaeger_2002_training,lee2023perspective,Allwood_2023_APL,Cucchi_2022_tutorial}. Training in RC schemes involves adjusting $\mathbf{W}_{\text{out}}$ using the reservoir states for a given set of input-output pairs, typically performed using regularised linear regression methods such as ridge regression~\cite{hilt_seegrist_1977}. This efficient training process contrasts with conventional deep neural networks (DNNs), which require fine-tuning interconnected node weights across multiple layers through iterative training processes such as backpropagation, often in the order of millions~\cite{Jaeger_2002_training}. 

\subsection{Design considerations of the reservoir}

When designing the RC network, one must consider the following key properties of a reservoir: nonlinearity, dimensional expansion, and the fading memory/echo state property~\cite{Cucchi_2022_tutorial,nakajima2021reservoir,love2023spatial,jaeger2001short,Roy_2007,Dambre_2012}.

The nonlinearity and dimensional expansion (e.g. more output nodes than input nodes) of the reservoir facilitate the mapping of input features to a high-dimensional feature space, which enables one to solve nonlinear problems via linear training methods. Fading memory or the echo-state property describes the reservoir's ability to respond strongly to recent past inputs while remaining unaffected by those from the distant past. This property makes RC particularly attractive for processing temporal data with transient dependencies, such as forecasting chaotic signals~\cite{Maass_NeuralComm2002,Jaeger_2004,Khovanov_2021,jaeger2001echo,jaeger2001short}.

It should be noted that there is typically a trade-off between these properties of nonlinearity (NL) and memory capacity (MC)~\cite{Dambre_2012}. Although it has been seen that including nonlinear and linear elements in a reservoir can improve performance, it is important to be aware that in a single reservoir, this often comes at the cost of reduced memory, and this trade-off should be kept in mind when designing a reservoir~\cite{Inubushi2017,stenning2023neuromorphic}. Recent approaches have shown that designing networks of interconnected reservoirs can overcome this trade-off, a particularly promising route to enhancing the computational power of reservoir computing \cite{gallicchio2017deep,gallicchio2017echo,manneschi2021exploiting,manneschi2024optimising}.

\

The nonlinearity and linear memory capacity of a given reservoir in \textit{PRCpy} are both calculated using the same methodology as Ref. \cite{love2023spatial}, which the following details. A linear estimator can be represented by:

\begin{equation}
\hat{y}(t) = \sum_{i} w_{i} u(t)
\end{equation}

where the weights, $w$, have been fitted using the training data inputs, $u(t)$ and outputs $y(t)$. The $R^{2}$ coefficient can be used to compare the estimated outputs, $\hat{y}(t)$ to the true outputs by evaluating:

\begin{equation}
R^{2}[\hat{y},y] = \frac{\text{cov}^{2}(\hat{y}(t),y(t))}{\sigma^{2}(\hat{y}(t))\sigma^{2}(y(t))}
\end{equation}

The variance is represented by $\sigma^{2}$ and the covariance is expressed as "$\text{cov}$". The nonlinearity can then be interpreted as:

\begin{equation}\label{eqn:nl}
\Phi^{\text{NL}}_{n} = 1 - R^{2}[\hat{y}_{n},y_{n}]
\end{equation}

While the memory capacity of the reservoir can be retrived by summing the $R^{2}$ coefficients for all delayed estimators, $\hat{u}_{n}(t-\tau)$, and inputs, $u(t-\tau)$, up to a threshold $k_{max}$:

\begin{equation}\label{eqn:mc}
\Phi^{\text{MC}}_{n} = \sum^{k_{max}}_{\tau=1} R^{2}[\hat{u}_{n}(t-\tau),u(t-\tau)]
\end{equation}

where $\tau$ is a time delay.

\section{Using \textit{PRCpy}}
\label{PRCpy_RC}

In this section, we introduce \textit{PRCpy}, a Python library that provides a high-level interface for implementing PRC. The program is designed to be flexible and extensible, allowing researchers to easily integrate it into their existing workflows with much room for customisability. It includes modules for data handling, preprocessing, model training, and evaluation. Moreover, two reservoir metric functions are provided, which can be used to analyse the intrinsic properties of a given PRC system.

We begin by installing the library and describing its key features through an example of using \textit{PRCpy} to solve a benchmark problem from real experimental data. The section concludes with a discussion of some of the library's advanced features and provides guidance on how to use them effectively.

\subsection{License and codebase}
The \textit{PRCpy} package is open source with \href{https://github.com/OSJL-py/PRCpy/blob/main/LICENSE}{MIT license}. The Git Repository is available at:\\ \url{https://github.com/OSJL-py/PRCpy}.

\subsection{Installation}
\textit{PRCpy} requires Python 3.9 or later and has been tested on a Windows device. It can be installed using pip or Poetry.

To install \textit{PRCpy} using pip, run:

\begin{lstlisting}[language=bash]
pip install prcpy
\end{lstlisting}

To install \textit{PRCpy} using Poetry, run:

\begin{lstlisting}[language=bash]
poetry add prcpy
\end{lstlisting}

It is recommended to use the latest release of \textit{PRCpy}, which can be obtained by running:

\begin{lstlisting}[language=bash]
PIP: pip install prcpy --upgrade
POERTY: poetry update prcpy
\end{lstlisting}

\subsection{General usage overview}
The general workflow for using \textit{PRCpy} consists of the following steps:

\begin{enumerate}
\item Define the data path.
\item Define the preprocessing parameters.
\item Create the RC pipeline.
\item Define the target and add it to the pipeline.
\item Define the model for training.
\item Define the RC parameters.
\item Run the RC pipeline.
\end{enumerate}

Here, we provide a brief overview of each step. More detailed explanations and examples are provided in Sec. \ref{PRCpy_examples}.

\subsubsection{Defining the data path}
The first step is to specify the directory containing the raw data files. The data file prefix parameter must also be specified (in all examples the prefix is set to "scan"). Example data files can be found in the \verb|examples/data_full| directory of the \textit{PRCpy} repository.

\subsubsection{Defining the preprocessing parameters}
Next, a dictionary containing the preprocessing parameters must be defined. These parameters specify how the raw data should be processed before being fed into the reservoir and a description of each parameter can be found in Table \ref{tab:PRCpyParams}.

\begin{table}[]
\caption{Description of available \textit{PRCpy} parameters for creating the reservoir matrix.}
\vspace{2mm}
\label{tab:PRCpyParams}
\begin{tabularx}{\columnwidth}{>{\raggedright}p{3cm}XX}
\toprule

\textbf{Parameter} & \textbf{Description} & \textbf{Notes} \\ \midrule

Xs & The name of the column containing the x-values (e.g., frequency). & Independent variable name. \\

\rowcolor{LightCyan}
Readouts & The name of the column containing the y-values (e.g., spectra). & Measured variable name. \\

delimiter & Specify measured data delimiter. & - \\

\rowcolor{LightCyan}
remove \textunderscore{} bg & Whether to remove the background from the data. & If True, will remove data from bg \textunderscore{} fname. \\

bg \textunderscore{} fname & Name of the background data file. & Note that the columns names must match Xs and Readouts. \\

\rowcolor{LightCyan}
smooth & Whether to apply smoothing to the data. & Applies a savgol \textunderscore{} filter. See \href{https://docs.scipy.org/doc/scipy/reference/generated/scipy.signal.savgol_filter.html}{savgol \textunderscore{} filter} for details. \\

smooth \textunderscore{} win & Smoothing filter window length. & Ensure smooth \textunderscore{} win is less than the number of measured data. \\

\rowcolor{LightCyan}
smooth \textunderscore{} rank & Smoothing filter polynomial order. & - \\

cut \textunderscore{} xs & Whether to slice the data array. & If True, the measure data (readouts) below x1 and above x2 will be removed when creating the reservoir. \\

\rowcolor{LightCyan}
x1 & Initial slicing data point. & - \\

x2 & Ending slicing data point. & - \\

\rowcolor{LightCyan}
normalize\_local & Whether to normalize each data channel separately. & If True, the measured data (readouts) will be normalized between 0 and 1. \\

normalize\_global & Whether to normalize all data channels. & - \\

\rowcolor{LightCyan}
sample & Whether to sample data. & If True, every 'sample \textunderscore{} rate' of the measured data (readouts) will be used when creating the reservoir. \\

sample \textunderscore{} rate & Frequency of sampling. & - \\

\rowcolor{LightCyan}
transpose & Transposes the RC dataframe. & This cannot be used simultaneously with other optional preprocessing parameters (as of v0.1.14).\\

\bottomrule
\end{tabularx}

\end{table}

\subsubsection{Creating the RC pipeline}
The RC pipeline encapsulates the entire workflow, from data loading to model evaluation. To create the pipeline object, pass the data directory path, file prefix and preprocessing parameters to the \verb|Pipeline| constructor, as shown in the example below:

\begin{lstlisting}[language=python]
from prcpy.RC import Pipeline

data_dir_path = "your/data/path"
prefix = "scan"
process_params = {
    "Xs": "independent_column": str,
    "Readouts": "readout_column": str,
    "delimiter": "\t": str,
    "remove_bg": True: bool,
    "bg_fname": "background_data.txt": str,
    "smooth": False: bool,
    "smooth_win": poly_window: int,
    "smooth_rank": poly_rank: int,
    "cut_xs": False: bool,
    "x1": x_limit_lower: int/float,
    "x2": x_limit_upper: int/float,
    "normalize_local": False: bool,
    "normalize_global": False: bool,
    "sample": True: bool,
    "sample_rate": adjacent_separation: int ,
    "transpose": False: bool }

rc_pipeline = Pipeline(data_dir_path, prefix, process_params)
\end{lstlisting}

\subsubsection{Defining the target}
The next step is to define the target signal. \textit{PRCpy} provides utility functions for generating common target signals, such as square waves and Mackey Glass time series.

For example, to generate a square wave target signal, one can use the \verb|generate_square_wave()| function:

\begin{lstlisting}[language=python]
from prcpy.Maths.Target_functions import generate_square_wave

length = rc_pipeline.get_df_length()
num_periods = 10

target_values = generate_square_wave(length, num_periods)
\end{lstlisting}

The target signal is then added to the RC pipeline using the \verb|define_target| method:

\begin{lstlisting}[language=python]
rc_pipeline.define_target(target_values)
\end{lstlisting}

\subsubsection{Defining the model}
\textit{PRCpy} provides a set of predefined models for training the reservoir, such as ridge regression and logistic regression. The model is defined by passing a dictionary which specifies its parameters.

For example, to define a ridge regression model, the \verb|define_Ridge| method can be called:

\begin{lstlisting}[language=python]
from prcpy.TrainingModels.RegressionModels import define_Ridge

model_params = {
    "alpha": 1e-3: int,
    "fit_intercept": True: bool,
    "copy_X": True: bool,
    "max_iter": None: int,
    "tol": 0.0001: float,
    "solver": "auto": str,
    "positive": False: bool,
    "random_state": None: int }

model = define_Ridge(model_params)
\end{lstlisting}

\subsubsection{Defining the RC parameters}
The RC parameters specify how the reservoir is trained and evaluated. The key parameters include:

\begin{itemize}
\item \verb|model|: The model to use for training.
\item \verb|tau|: The delay between the input and target signals. This value can be specified to predict \verb|tau| steps in to the future for forecasting tasks while \verb|tau| must be set to zero for transformation tasks. 
\item \verb|test_size|: The fraction of the data to use for testing.
\item \verb|error_type|: The type of error to use for evaluation (e.g., MSE).
\end{itemize}

Relevant RC parameters for training and performance evaluations are specified in a dictionary format:

\begin{lstlisting}[language=python]
rc_params = {
    "model": model,
    "tau": 0: int,
    "test_size": 0.3: float,
    "error_type": "MSE": {"MSE", "MAE"} }
\end{lstlisting}

\subsubsection{Running the RC pipeline}
Finally, the RC pipeline can be executed by calling the \verb|run| method and passing the \verb|rc_params| dictionary:

\begin{lstlisting}[language=python]
rc_pipeline.run(rc_params)
\end{lstlisting}

This will automatically load the data, perform preprocessing, train the model, and evaluate its performance on the test set.

\subsubsection{Getting the results}
\label{sec_getting_results}
After running the RC pipeline, the results can be obtained by using the \verb|get_rc_results| method:

\begin{lstlisting}[language=python]
results = rc_pipeline.get_rc_results()
\end{lstlisting}

The results are returned as a dictionary containing:

\begin{lstlisting}[language=python]
results_df = {
    "train": {
        "x_train": x_train,
        "y_train": y_train,
        "train_pred": train_predictions},
        
    "test": {
        "x_test": x_test,
        "y_test": y_test,
        "test_pred": test_predictions},
        
    "error": {
        "train_error": train_error: float,
        "test_error": test_error: float}   }
\end{lstlisting}

The nonlinearity and memory capacity can also be evaluated using the following procedure.
First define the input signal used to generate the readouts being evaluated:

\begin{lstlisting}[language=python]
rc_pipeline.define_input(input_values)
\end{lstlisting}

Then use the following functions to calculate either nonlinearity (NL) or linear memory capacity (LMC):

\begin{lstlisting}[language=python]
nl = rc_pipeline.get_non_linearity()

total_mc, mc_list = rc_pipeline.get_linear_memory_capacity(kmax=25, remove_auto_correlation=True)
\end{lstlisting}

The first command produces the average NL score for the entire reservoir ranging from 0 for completely linear to 1 for completely nonlinear by utilising Eqn. \ref{eqn:nl}. 
The second provides two outputs: total LMC and the LMC list. The latter provides a score between 0 and 1 for the ability of a set of readouts to predict a previous input \(k\) steps ago up to the user defined \(k_{max}\) value (\(k_{max}\) is set to 25 by default). The total LMC is the summation of the LMC list (Eqn. \ref{eqn:mc}). The "remove\_auto\_correlation" parameter subtracts the correlation at each $k$ value of the input signal with itself which is set to False by default. This parameter can aid the user in removing the intrinsic memory imparted from a quasi-periodic input signal, such as Mackey Glass, but can lead to unmeaningful negative LMC scores if an inappropriate $k_{max}$ value is used.

\section{Tutorial examples}
\label{PRCpy_examples}

Here, we provide a step-by-step guide for using \textit{PRCpy} on three different reservoir systems (the code files can be found in the \verb|examples/tutorials| directory of the \textit{PRCpy} repository). In Sec.~\ref{example_transformation}, we construct a reservoir using a resistor and a red LED by time multiplexing to perform transformation tasks. In Sec.~\ref{example_tr_for} we add a capacitor to the aforementioned circuit to demonstrate the importance of memory in a number of tasks, and utilise the reservoir metrics included in the \textit{PRCpy} package to compare these two electronic reservoirs. Section~\ref{example_forecasting} demonstrates PRC with a more complex system involving a nontrivial magnetic material with frequency-multiplexing.

\subsection{Diode PRC}
\label{example_transformation}

\begin{figure}[h]
\centering
\includegraphics[width=1\linewidth]{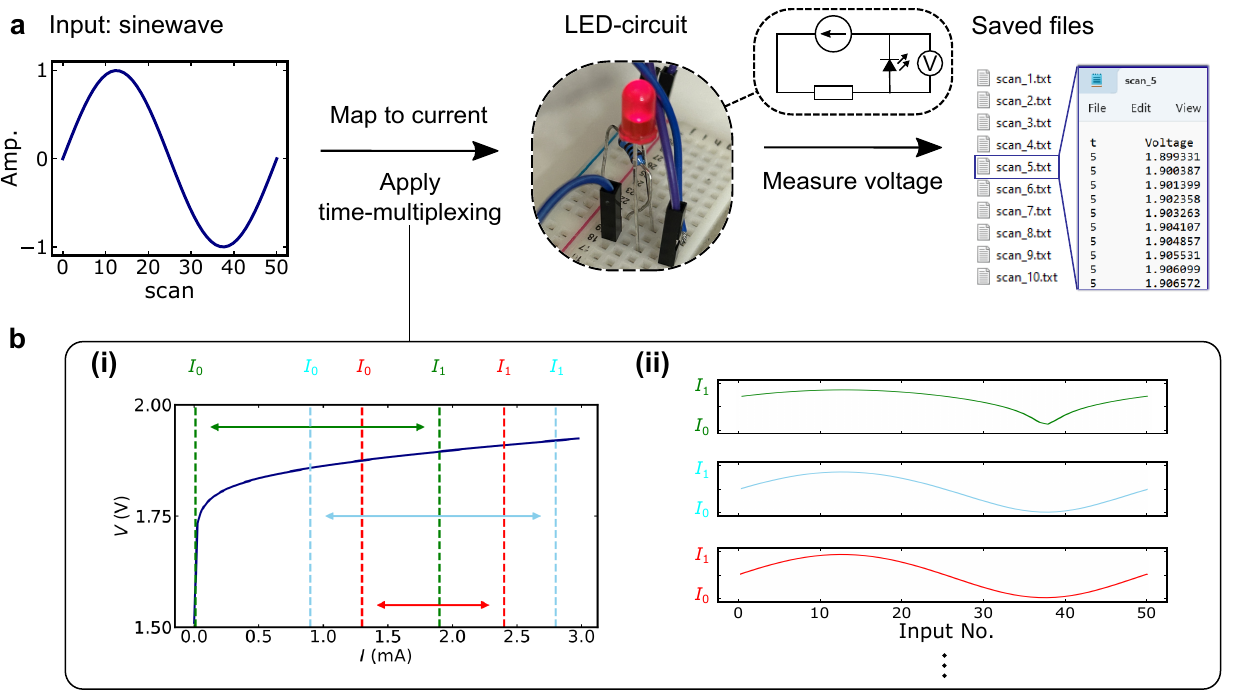}
\caption{\textbf{Creating the reservoir layer.} \textbf{a}, A pre-defined input signal ($u(t)=\rm{sin}(\textit{t})$) is mapped onto the current and applied to a simple LED circuit as shown in the middle panel. For each current, voltage is measured and is individually saved to a file. Here, ``scan” refers to the next measurement number. \textbf{b}, Time-multiplexing is employed to increase the dimensionality of the reservoir for single-input-single-output systems as such. Three example windows are shown with randomly selected bounds on the \textit{I-V} curve (\textbf{b (i)}). The corresponding mapped currents (\textbf{b (ii)}) are subsequently applied.}
\label{Fig_creating_reservoir}
\end{figure}

Figure~\subfigref{Fig_creating_reservoir}{a} depicts the experiment schematic for constructing a diode reservoir. Note that such implementation is constructed for the sole purpose of demonstrating \textit{PRCpy}. Here, a standard red LED is connected in series with a resistor, and the voltage, $V$, is measured as a function of the input current, $I$. The \textit{I-V} characteristics of an LED are nonlinear as shown in Fig.~\subfigref{Fig_creating_reservoir}{b(i)}. The input to the reservoir, $\mathbf{x}_{\text{in}}(t)$ is a 10-period sinewave consisting of 50 points per period.

One method to increase the dimensionality of the reservoir is to utilise the time-multiplexing or `virtual node' technique as introduced by Appeltant et al.~\cite{appeltant2011information}. This scheme involves randomly choosing $N$ pairs of bounds on the \textit{I-V} curve, which represent $N$ windows. Three examples of these windows are shown by the dashed vertical lines in Fig.~\subfigref{Fig_creating_reservoir}{b(i)}. Subsequently, $\mathbf{x}_{\text{in}}(t)$ is mapped onto $I$ for each window, $n_{i}$ at a given time, $t$ (Fig.~\subfigref{Fig_creating_reservoir}{b(ii)}). This approach allows the efficient utilisation of a single-input-single-output system in creating an $N$-dimensional reservoir. At each applied value of $I$, a separate file is saved with the "scan" prefix indicating the measurement number.

\newpage

\textit{PRCpy} allows one to create a \textbf{customised reservoir matrix}. No preprocessing steps are activated by default and can be left empty unless specified. The code below shows the necessary steps for creating and viewing the reservoir matrix. Note that all file names must match the specified prefix string (for this example we have set prefix="scan", see right panel of Fig.~\subfigref{Fig_creating_reservoir}{a}).

\begin{lstlisting}[language=python]
""" setting up PRCpy """

## essential module
from prcpy.RC import Pipeline

## optional modules (used for this example)
from prcpy.TrainingModels.RegressionModels import define_Ridge
from prcpy.Maths.Target_functions import generate_square_wave
import numpy as np
import matplotlib.pyplot as plt

## specify data directory path
data_dir_path = r"examples/data_100/sine_mapping/diode"
prefix = "scan"

## define processing parameters
process_params = {
    "Xs": "t",
    "Readouts": "Voltage",
    "delimiter": "\t",
    "remove_bg": False,
    "smooth": False,
    "cut_xs": False,
    "sample": False,
    "normalize_local": False,
    "normalize_global": False,
    "transpose": True
}

## create a pipeline object, encapsulating all RC features.
rc_pipeline = Pipeline(data_dir_path, prefix, process_params)

## view the reservoir matrix
reservoir_df = rc_pipeline.rc_data.rc_df
print(reservoir_df)

>>          r0        r1        r2        ...  r498      r499      r500
>> Scan1    1.787651  1.826460  1.846522  ...  1.902053  1.910276  1.918088
>> Scan2    1.791340  1.827726  1.847296  ...  1.902375  1.910583  1.918371
>> Scan3    1.794408  1.828896  1.848057  ...  1.902732  1.910921  1.918700
>> Scan4    1.797150  1.830044  1.848818  ...  1.903122  1.911282  1.919042
>> ...           ...       ...       ...  ...       ...       ...       ...
>> Scan497  1.768355  1.819889  1.841874  ...  1.899318  1.907673  1.915581
>> Scan498  1.773524  1.821212  1.842721  ...  1.899721  1.908047  1.915943      
>> Scan499  1.778505  1.822643  1.843633  ...  1.900164  1.908474  1.916348      
>> Scan500  1.782793  1.824015  1.844522  ...  1.900621  1.908901  1.916750            
>> 
>> [500 rows x 501 columns]


\end{lstlisting}

Next, the target function must be defined and added to the reservoir dataframe. Targets can be generated by calling the functions defined in \texttt{prcpy.Maths.Target\_functions} or can be user-generated. To perform transformation tasks, one must ensure the period and the length of the input and the target are matched. This is automatically handled by \textit{PRCpy}. See the code below:

\begin{lstlisting}[language=python]
""" target generation (transformation) """

## define number of periods and length of the input signal
length = rc_pipeline.get_df_length()
num_periods = 10

## define target
target_values = generate_square_wave(length, num_periods)

## add target to the reservoir matrix
rc_pipeline.define_target(target_values)
\end{lstlisting}

The RC training parameters must be specified before running the model. This includes the model definition, test size (percentage as a decimal), and the error type (supports: ``MSE": Mean Square Error (MSE) and "MAE": Mean Absolute Error) for performance evaluation. By default, \textit{PRCpy} provides three simple regression models: ridge, linear and logistic regression, utilising the \texttt{sklearn} module. Custom models can also be applied and specified by the RC parameters.

\begin{lstlisting}[language=python]
""" training parameters definition (transformation) """

## define the training model. We will choose the ridge regression.
model_params = {
    "alpha": 1e-6,
    "fit_intercept": True,
    "copy_X": True,
    "max_iter": None,
    "tol": 0.0001,
    "solver": "auto",
    "positive": False,
    "random_state": None,
}
model = define_Ridge(model_params)

## define the RC parameters
rc_params = {
    "model": model,
    "tau": 0, # transformation task
    "test_size": 0.3,
    "error_type": "MSE"
}

\end{lstlisting}

\begin{figure}[h]
\centering
\includegraphics[width=1\linewidth]{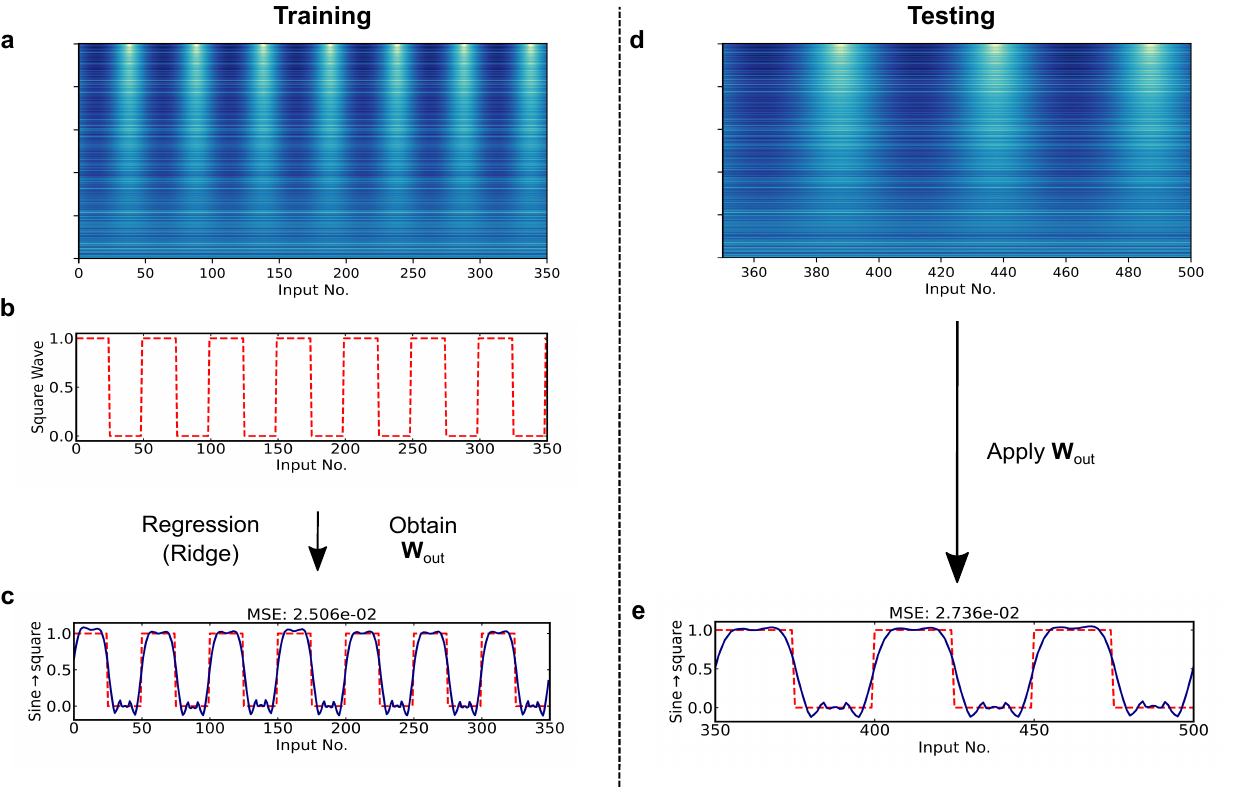}
\caption{\textbf{Transformation RC example.} \textbf{a(d)}, A visualisation of the reservoir matrix for the training (testing) data. \textbf{b}, Defined target signal ($y(t)=\rm{square}(\textit{t})$). Ridge regression is applied to the training set of the reservoir matrix and their corresponding target values to obtain an optimised weight matrix. \textbf{c(e)}, Performance of training – the weights are applied to the training (testing) reservoir matrix, which is shown by a blue line, and compared with the target signal (dashed red line).}
\label{Fig_performingDiodeRC}
\end{figure}

Once the RC parameters are defined, the model can be passed to the \texttt{run} function. This will split the reservoir matrix and its targets into training (learning) and testing (unseen) datasets in a ratio according to the user-specified value of "test\_size". The obtained weights from the regression are then applied to the testing dataset to perform the computation as shown in Figs.~\subfigref{Fig_performingDiodeRC}{a-e}.

\begin{lstlisting}[language=python]
""" perform RC (transformation) """

## run the pipeline
rc_pipeline.run(rc_params)

\end{lstlisting}

Finally, the results of the computation can be accessed by the \texttt{get\_rc\_results()} function. This returns the relevant data frames used for computation, prediction and error analysis. See Sec.~\ref{sec_getting_results} for details. 

\begin{lstlisting}[language=python]
""" obtaining PRC results """

results = rc_pipeline.get_rc_results()

## it may be beneficial to attribute each result into its own variable.

# targets
train_ys = results["train"]["y_train"]
test_ys = results["test"]["y_test"]

# predictions
train_preds = results["train"]["train_pred"]
test_preds = results["test"]["test_pred"]

# errors
train_MSE = results["error"]["train_error"]
test_MSE = results["error"]["test_error"]

# x-axis 
train_ts = np.arange(train_ys.shape[0])
test_ts = np.arange(test_ys.shape[0])

\end{lstlisting}

Results of PRC performance can be evaluated by viewing the magnitude of the error or plotting the training and testing data with their target values, as seen in Figs.~\subfigref{Fig_performingDiodeRC}{c\&e}.

\begin{lstlisting}[language=python]
""" Evaluating the PRC performance """

## error
print(f"Training MSE = {train_MSE, '0.3e'}")
print(f"Testing MSE = {test_MSE, '0.3e'}")

>>> Training MSE = 2.506e-02
>>> Testing MSE = 2.736e-02
\end{lstlisting}

\begin{lstlisting}[language=python]
""" Evaluating the PRC performance """

## plotting
plt.plot(train_ts, train_ys, label="Train", color="red", ls="--")
plt.plot(train_ts, train_preds, label="Train predict", color="navy")
plt.show()

\end{lstlisting}

It should be noted that while we refer to this diode circuit as a reservoir, it infact does not fulfil the echo state property \cite{jaeger2001echo}, and therefore is not conventionally defined as a reservoir. A more concise description would be that we have used a ridge regression layer on the diode's readout. Despite this technicality, the presence of nonlinearity does enable it to perform tasks with no memory requirement.

\subsection{Capacitor \& diode PRC}
\label{example_tr_for}

\begin{figure}[h]
\centering
\includegraphics[width=1\linewidth]{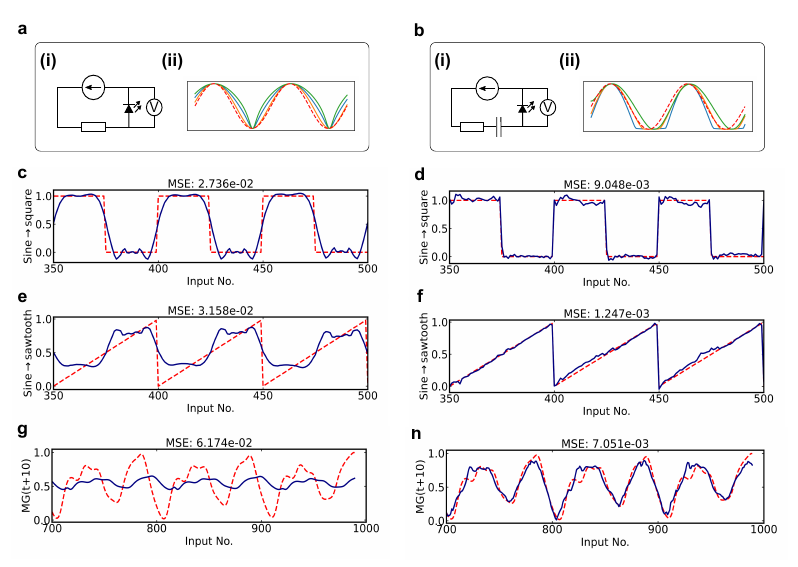}
\caption{\textbf{Comparison between diode circuit and capacitor \& diode circuit RC performances. a(i)}, A diode circuit diagram. \textbf{a(ii)}, Three example line profiles from the diode reservoir. \textbf{b(i)}, A capacitor \& diode circuit diagram. \textbf{b(ii)}, Three example line profiles from the capacitor \& diode reservoir. \textbf{c \& d}, Comparison between sine to square wave and (\textbf{e \& f}), sine to sawtooth wave transformation testing performances for both circuits. \textbf{g \& h}, Comparison between Mackey Glass forecasting testing performances for both circuits.}
\label{Fig_diode_capacitor_comp}
\end{figure}

This example uses the same circuit as in Sec. \ref{example_transformation} with the addition of a capacitor as shown in Fig~\subfigref{Fig_diode_capacitor_comp}{b(i)}. The capacitor here, along with the resistor, is indented to cause voltage decay that provides memory of previous inputs. This memory effect is evident when comparing the output line profiles to the sine input for the diode only (Fig~\subfigref{Fig_diode_capacitor_comp}{a(ii)}) and the capacitor \& diode reservoir (Fig~\subfigref{Fig_diode_capacitor_comp}{b(ii)}). The output depends only on the input in Fig~\subfigref{Fig_diode_capacitor_comp}{a(ii)} while the output depends both on the input and position on the sine wave for Fig~\subfigref{Fig_diode_capacitor_comp}{b(ii)}. This memory is what allows the circuit to perform well for forecasting tasks, such as predicting the Mackey Glass time series.

The contrast in performance for the three tasks shown in Fig \ref{Fig_diode_capacitor_comp} between both circuits can be explained by the reservoir metrics (Table \ref{metric_table}). The MSEs for both transformation tasks: sine to square wave (Fig~\subfigref{Fig_diode_capacitor_comp}{c\&d}) and sine to sawtooth wave (Fig~\subfigref{Fig_diode_capacitor_comp}{e\&f}), differ between the two circuits. While the sine to square task is purely nonlinear, the sawtooth does require some memory as the transformation is not just a one-to-one mapping from the input to the output. Information about the position on the sine wave is required. The capacitor \& diode's superior nonlinearity score of $0.151$ exhibits an improved MSE of $9.048 \times 10^{-3}$ compared to the diode's MSE of $2.736 \times 10^{-2}$. The sine to sawtooth tasks only requires a small amount of memory capacity to effectively transform the input data, and as such the capacitor \& diode far outperform the diode only reservoir. Figure~\subfigref{Fig_diode_capacitor_comp}{g} shows how poorly the diode performs when predicting a Mackey Glass signal. While the MSE score is on the order of $10^{-2}$, it should be noted that a straight line at the $\text{MG}(t+10)=0.5$ produces an MSE of $\sim$ $7 \times 10^{-2}$. On the other hand, the capacitor \& diode reservoir performs significantly better with an MSE of $1.247 \times 10^{-3}$. This is explained by the greater MC score of 5.201 compared to the diode only score of $1.976$.

The MC score presented in this work of $1.976$ appears to be rather high for the diode only reservoir that ostensibly has no memory. This reveals an issue with using MC to characterise the memory of the reservoir. For our calculation, we used the Mackey Glass input as it best demonstrated the memory that exists within our reservoir. This input, however, has some self-correlation that allows it to obtain a non-zero MC score on itself. As such, we include an option to remove this contribution ("remove\_auto\_correlation"=True), but this feature is not comprehensive as discussed in Sec. \ref{sec_getting_results}. Despite this, MC is useful when comparing reservoirs using the same input method, as we show in Table \ref{metric_table}.

\begin{table}
\centering
\caption{Comparison between reservoir metrics for diode and capacitor \& diode circuits.}
\vspace{2mm}
\begin{tabular}{ |c|c|c| } 
\hline
\textbf{Reservoir} & \textbf{Nonlinearity} & \textbf{Linear memory capacity}\\
\hline
Diode & 0.060 & 1.976 \\
\rowcolor{LightCyan}
Capacitor \& Diode & 0.151 & 5.201 \\ 
\hline
\end{tabular}
\label{metric_table}
\end{table}

\newpage

The nonlinearity and linear memory capacity reservoir metrics can be obtained by running:

\begin{lstlisting}[language=python]
""" Evaluating reservoir properties """

mg_path = "examples/data_full/chaos/mackey_glass_t17.npy"
input_values = get_npy_data(mg_path, norm=True)

## metrics
rc_pipeline.define_input(input_values[:1000])

nl = rc_pipeline.get_non_linearity()
lmc = rc_pipeline.get_linear_memory_capacity(remove_auto_correlation=True, kmax = 12)[0]

print(f"NL = {nl}")
print(f"LMC = {lmc}")

>>> NL = 0.1507705132191488
>>> LMC = 5.200838123059376
\end{lstlisting}

\subsection{Magnetic PRC}
\label{example_forecasting}

In this example, a magnetic system (Cu$_2$OSeO$_3$) is realised as a physical reservoir as shown by prior works in Ref.~[\citenum{lee2024task}]. This system has a large memory capacity originating from nontrivial spin-textures (magnetic skyrmions) and is suitable for future forecasting tasks. The input signal to the reservoir is defined by a Mackey Glass signal~\cite{Mackey_1977}, tuned numerically for chaotic behaviour (see Ref.~[\citenum{lee2024task}] for details). The target signal is the same as the input, however, with a 10-future step offset in the x-axis. This is equivalent to using the current data $t_i$ to predict $t_{i+10}$. 

Moreover, a different multiplexing technique to the electronic PRC examples is employed. Here, frequency-multiplexing is utilised, and for every input (external magnetic field), the power absorption from a range of frequencies is measured using specialised experimental equipment (vector network analyser). See Ref.~[\citenum{lee2024task}] for full details. For further reading on skyrmion-based reservoir computing, see also a comprehensive review in Ref.~\cite{lee2023perspective} and references therein.

\begin{figure}[h]
\centering
\includegraphics[width=1\linewidth]{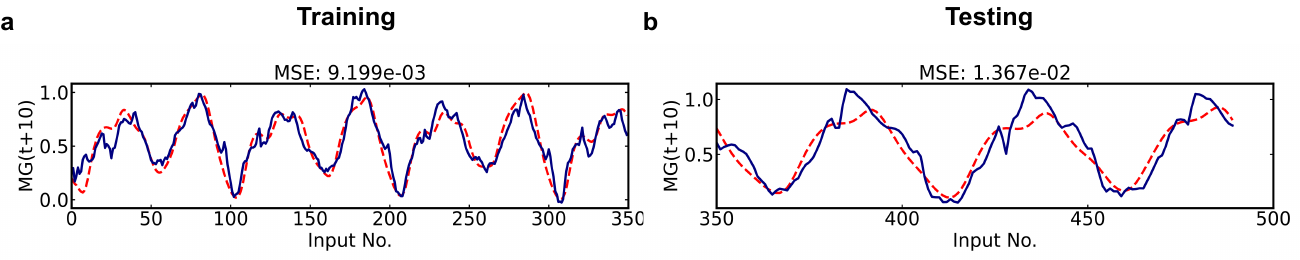}
\caption{\textbf{Forecasting RC example.} Forecasting of a Mackey Glass signal for 10 future steps using a magnetic (skyrmion) reservoir.}
\label{skyrmionRC}
\end{figure}

The code for performing the computation with \textit{PRCpy} is presented below. The final result of the prediction is shown in Fig.~\ref{skyrmionRC}.

\begin{lstlisting}[language=python]
""" Magnetic reservoir (forecasting)"""

from prcpy.RC import Pipeline
from prcpy.TrainingModels.RegressionModels import define_Ridge
from prcpy.Maths.Target_functions import get_npy_data
from prcpy.Maths.Maths_functions import normalize_list

data_dir_path = r"examples/data_full/mg_mapping/Cu2OSeO3/skyrmion"
prefix = "scan"
mg_path = "examples/data_full/chaos/mackey_glass_t17.npy"

process_params = {
    "Xs": "Frequency",
    "Readouts": "Spectra",
    "delimiter": ",",
    "remove_bg": True, 
    "bg_fname": "BG_450mT_1_0to6_0GHz_4K.txt", 
    "sample": True, 
    "sample_rate": 13,
    "normalize_local": False,
    "normalize_global": False,
    "transpose": False
}

rc_pipeline = Pipeline(data_dir_path, prefix, process_params)

target_values = normalize_list(get_npy_data(mg_path))
rc_pipeline.define_target(target_values)

model_params = {
    "alpha": 1e-3,
    "fit_intercept": True,
    "copy_X": True,
    "max_iter": None,
    "tol": 0.0001,
    "solver": "auto",
    "positive": False,
    "random_state": None,
}

model = define_Ridge(model_params)

rc_params = {
    "model": model,
    "tau": 10, # forecasting of 10 steps into the future
    "test_size": 0.3,
    "error_type": "MSE"
}

rc_pipeline.run(rc_params)

\end{lstlisting}

\section{Conclusion and outlook}
\label{conclusion}

This manuscript has covered a brief introduction to PRC and its practical use cases utilising the \textit{PRCpy} package on three physical systems. While excellent literature and reviews have already presented thorough workings of (P)RC, its practical implementations remained nontrivial, bottlenecked by an adequate understanding of programming knowledge requirements. 

Here, we provide tools to allow more researchers from diverse communities to participate in PRC research with their unique systems. Despite numerous proposals and demonstrations, the PRC field requires more steps before sparking its route for commercialisation. Future research into reservoir systems needs to consider their viability for mass production and scalability. The practical implementation in terms of the energy consumption, footprint and operating conditions also needs consideration to enable the application of PRC to edge computing.

We understand that the research is no longer a singular study. Researchers from different backgrounds are encouraged to collaborate within multidisciplinary areas not limited to physics, electrical engineering, or computer science, to explore possible paths for designing and engineering future energy-efficient computing systems. Given the range of different physical systems that can be used in the implementation of PRC, it stands out as a field particularly opportune for this kind of collaboration. As previously stated, the \textit{PRCpy} package remains open source, and we encourage all researchers to contribute to making the project more widely available.
\section*{Acknowledgments}
\label{acknowledgments}
The authors thank Dr. Matthias Sitte for providing insightful advice to improve the quality and standards of the codebase. 

\bibliography{references}

\end{document}